\documentclass[prl,twocolumn,groupedaddress,showpacs,nofootinbib]{revtex4}
\usepackage{graphicx}
\usepackage{dcolumn}
\usepackage{bm}
\usepackage{amssymb}
\usepackage{mathrsfs}
\usepackage{amsmath}
\usepackage{epsfig}
\usepackage[dvips]{color}
\usepackage{hhline}

\newcommand{\beq}{\begin{equation}}
\newcommand{\eeq}{\end{equation}}
\newcommand{\bq}{\begin{equation}}
\newcommand{\eq}{\end{equation}}
\newcommand{\ba}{\begin{array}}
\newcommand{\ea}{\end{array}}
\newcommand{\beqa}{\begin{eqnarray}}
\newcommand{\eeqa}{\end{eqnarray}}
\newcommand{\beqs}{\begin{subequations}}
\newcommand{\eeqs}{\end{subequations}}

\def\f{\frac}
\def\nn{\nonumber}

\begin{document}

\title{
Double Type-II Seesaw, Baryon Asymmetry and Dark Matter for Cosmic
$\boldsymbol{e^\pm}$ Excesses }

\author{Pei-Hong Gu$^{1}_{}$}
\email{pgu@ictp.it}

\author{Hong-Jian He$^{2}_{}$}

\email{hjhe@tsinghua.edu.cn}

\author{Utpal Sarkar$^{3,4}_{}$}
\email{utpal@prl.res.in}

\author{Xinmin Zhang$^{5}_{}$}
\email{xmzhang@ihep.ac.cn}

\affiliation{$^{1}_{}$The Abdus Salam International Centre for
Theoretical Physics, Strada Costiera 11, 34014 Trieste, Italy\\
$^{2}$Center for High Energy Physics, Tsinghua University, Beijing 100084, China\\
$^{3}_{}$Physical Research Laboratory, Ahmedabad 380009, India\\
$^{4}_{}$Department of Physics and McDonnell Center for the Space
Sciences, Washington University, St.\ Louis, MO 63130, USA\\
$^{5}_{}$Institute of High Energy Physics, Chinese Academy of
Sciences, Beijing 100049, China}

\begin{abstract}
\vspace*{-4mm}

We construct a new realization of type-II seesaw for neutrino masses
and baryon asymmetry by extending the standard model with one light
and two heavy singlet scalars besides one Higgs triplet. The heavy
singlets pick up small vacuum expectation values to give a
suppressed trilinear coupling between the triplet and doublet Higgs
bosons after the light singlet drives the spontaneous breaking of
lepton number. The Higgs triplet can thus remain light and be
accessible at the LHC. The lepton number conserving decays of the
heavy singlets can generate a lepton asymmetry stored in the Higgs
triplet to account for the matter-antimatter asymmetry in the
Universe. We further introduce stable gauge bosons from a hidden sector,
which obtain masses and annihilate into the Higgs triplet after
spontaneous breaking of the associated non-Abelian gauge symmetry.
With Breit-Wigner enhancement, the stable gauge bosons can simultaneously
explain the relic density of dark matter and the cosmic positron/electron
excesses.

\vspace*{-4mm}
\end{abstract}


\maketitle

%\vspace*{2mm}

\section{I. Introduction}

Observations of solar, atmospheric, reactor and accelerator
neutrino oscillations have strongly pointed to
tiny but nonzero neutrino masses \cite{ms2006}.
The smallness of neutrino masses is naturally explained by
the seesaw \cite{minkowski1977} extension of the standard model (SM).
In the seesaw scenario, the observed baryon asymmetry in the Universe
can be generated through the leptogenesis \cite{fy1986,lpy1986}.
The type-II seesaw models \cite{mw1980} generically contain
a trilinear coupling between the triplet and doublet Higgs bosons, which is
a unique source for the lepton number violation. This interaction is the key
ingredient to realize leptogenesis in the type-II seesaw \cite{ms1998}.
If this lepton number violation is very weak, a lepton asymmetry stored
in the Higgs triplet can survive and thus be transferred to the lepton
doublets.  Subsequently the sphaleron process \cite{krs1985}
can partially convert this lepton asymmetry to the observed
baryon asymmetry in the Universe.
When the type-II seesaw is implemented in models of large extra
dimensions with lepton number breaking in a distant brane, it can
lead to interesting collider phenomenology by predicting a light
Higgs triplet with significant couplings to leptons \cite{mrs2000}.

In this paper, we construct a new double type-II seesaw scenario to
simultaneously generate a suppressed coupling of the Higgs triplet
to the Higgs doublet and a lepton asymmetry stored in the Higgs
triplet. This is realized by introducing two heavy and one light
singlet scalars to the type-II seesaw model, where the Higgs triplet
has a TeV-scale mass and its Yukawa couplings to the lepton doublets
are naturally ${\cal O}(1)$, so it is within the discovery reach of
the LHC. When the light singlet develops a vacuum expectation value
(VEV), the lepton number will be spontaneously broken at the TeV
scale. The heavy singlets will then pick up small VEVs to give the
suppressed coupling between the Higgs triplet and doublet. These heavy
singlets are also responsible for generating the lepton asymmetry
stored in the Higgs triplet through their out-of-equilibrium decays
which violate CP but conserve lepton number.

On the other hand, strong evidences for the non-baryonic dark matter
relic abundance\,\cite{bhs2005} require supplementing additional new
ingredients to the existing theory. The dark matter may also be
responsible for the positron/electron excesses in the cosmic rays as
observed by ATIC\,\cite{chang2008}, PPB-BETS\,\cite{torii2008},
PAMELA\,\cite{adriani2008}, HESS\,\cite{aharonian2008} and
Fermi/LAT\,\cite{fermi2009} collaborations. This indicates that the
dark matter should mostly annihilate or decay into leptons with
large cross section or long lifetime. Such type of dark matter may
have special relation with the neutrino
mass-generation\,\cite{bglz2009}. In this study, we extend the
double type-II seesaw model for the neutrino masses to explain the
relic abundance of dark matter \cite{komatsu2008} as well as the
cosmic positron/electron excesses. In a bosonic hidden sector, the
vector bosons associated with a non-Abelian gauge group can be
stable since they are forbidden to mix with the gauge bosons of the
SM due to the non-Abelian character. Such type of hidden vector
bosons can explain the relic density of dark matter in the Universe
\cite{hambye2008}. In the present construction, these hidden vector
bosons will annihilate into the Higgs triplet in our double type-II
seesaw scenario and the Higgs triplet dominantly decays into lepton
pairs. The annihilation process of the hidden vector bosons into the
Higgs triplets invokes an $s$-channel exchange of a Higgs boson
(which is also responsible for the mass generation of dark matter).
With the Breit-Wigner resonant enhancement
\cite{RS-GS,ckrs2008,imy2008,gw2009}, the cross section could be
large enough to account for the cosmic-ray positron/electron excess
without introducing any other additional boost factor. In this way
the hidden vector bosons can naturally serve as a new candidate of
leptonic dark matter \cite{bglz2009,fp2008,gos2009}.

\section{II. Double Type-II Seesaw}

The type-II seesaw mechanism is realized by extending the SM
with a Higgs triplet. Under the SM gauge group, this Higgs
triplet is allowed to have a Yukawa coupling with the lepton doublets,
\begin{eqnarray}
\label{yukawa} \mathcal{L_{\textrm{Y}}^{}} &\supset&
-\frac{1}{2}f\,\overline{\psi_{L}^{c}}
i\tau_2^{}\xi\psi_{L}^{}+\textrm{h.c.}\,,
\end{eqnarray}
and a trilinear interaction with the Higgs doublet in the scalar
potential,
\begin{eqnarray}
\label{trilinear} V &\supset&
-\mu_0^{}\,\phi_{}^{T}i\tau_2^{}\xi\phi+\textrm{h.c.}\,,
\end{eqnarray}
where $\psi_{L}^{}$ and $\phi$ denote the lepton and Higgs doublets,
respectively, while $\xi$ is the Higgs triplet. In the presence of
trilinear interaction (\ref{trilinear}), the Higgs triplet $\xi$ can
develop a small VEV once the Higgs doublet $\phi$ acquires its VEV
to break the electroweak symmetry. Such a small VEV of $\xi$ will
thus naturally generate tiny neutrino masses through the Yukawa
coupling (\ref{yukawa}). Conventionally, we assign the Higgs triplet
$\,\xi\, $ a lepton number \,$L=-2$\, as the lepton doublet $\psi_{L}^{}$
has \,$L=+1$\,.\, In consequence, the Yukawa interaction
(\ref{yukawa}) is lepton-number conserving, while the trilinear
interaction (\ref{trilinear}) softly and explicitly breaks the
lepton number because of $\,L=0\,$ for the Higgs doublet $\phi$\,.

To explain the origin of the lepton number violation in the
type-II seesaw model, it is desirable to naturally start with a lepton
number conserving Lagrangian and then break it spontaneously.
This can be achieved by introducing a
singlet scalar with a global lepton number
$\,L=+2\,$,\, similar to the singlet majoron model \cite{cmp1980}.
Then the cubic coupling in the lepton number violating interaction
(\ref{trilinear}) should be proportional to the breaking scale of
lepton number.

In the present study, we propose a more attractive scenario by
extending the type-II seesaw model with heavy and light singlet
scalars. For simplicity, we will not give the complete scalar
potential, instead we write the relevant part for our analysis as
follows,
\begin{eqnarray}
\label{potential}
V &\supset&
-m_{1}^{2}\left(\sigma^{\dagger}_{}\sigma\right)
+\lambda_{1}^{}\left(\sigma^{\dagger}_{}\sigma\right)^{2}_{}
-m_{2}^{2}\left(\phi^{\dagger}_{}\phi\right)+\lambda_{2}^{}
\left(\phi^{\dagger}_{}\phi\right)^{2}_{}
\nonumber\\[2mm]
&&
+M_\chi^2\left(\chi^\dag_{}\chi\right)
-\left(\mu\chi\sigma^2_{}+\textrm{h.c.}\right)
\nonumber\\[2mm]
&& +m_\xi^2\textrm{Tr}\left(\xi^\dagger_{}\xi\right)
-\left(\kappa\chi\phi^{T}_{}i\tau_2^{}\xi\phi+ \textrm{h.c.}\right),
\end{eqnarray}
where $\sigma$ and $\chi$ are the light and heavy singlets, respectively.
For the lepton number conservation in the scalar potential (\ref{potential}),
we have assigned $\,L=+2\,$ for $\chi$ and $\,L=-1\,$ for $\sigma$\,.
A summary of the relevant quantum number arrangement of our model is given
in Table.\,1.

\begin{table}
\begin{center}
\caption{Summary of relevant quantum number assignments in the present
 model, where the new gauge boson $X^a_\mu$ and Higgs doublet $\eta$ under
 the hidden gauge group $SU(2)_h$ will be explicitly defined in Sec.\,IV.}
 \vspace*{1.5mm}
\begin{tabular}{c||cccc}
 \hline\hline
 &&&& \\[-2mm]
  \quad & ~\,$SU(2)_{L}^{}$ \quad & $~~U(1)_{Y}^{}$ \quad &
             $U(1)_{\textrm{lepton}}^{}$ & $SU(2)_h$~~~
  \\[1.5mm]
 ~~Fields~~~ & ($I_{L3}$)  & ($Y$) &  ($L^{\#}$)  & ($I_{h3}$)~~~
  \\[1.5mm]
  \hline
 &&&& \\[-2mm]
  ~~$\psi_{L}^{}$  \quad\quad &     ~~\,$\f{1}{2}$    \quad\quad & $  -\frac{1}{2}
  \quad $ & $1$  & $0$~~~
  \\[1.5mm]
  ~~$\phi$         \quad\quad &     ~~\,$\f{1}{2}$    \quad\quad & $  -\frac{1}{2}
  \quad $ & ~$0$~  & $0$~~~
  \\[1.5mm]
  ~~$\xi$   \quad\quad &     ~~\,$1$    \quad\quad & $      ~~\,$1$
  \quad $ & $-2$~~  & $0$~~~
  \\[1.5mm]
  ~~~$\chi$         \quad\quad &     ~~$0$    \quad\quad &      $0$
  &  $2$  &  $0$~~~
  \\[1.5mm]
  ~~~$\sigma$       \quad\quad &     ~~$0$    \quad\quad &      $0$
  &  $-1$  &  $0$~~~
  \\[1.5mm]
  ~~~$\eta$         \quad\quad &     ~~$0$    \quad\quad & $0$ & $0$ & $\f{1}{2}$~~~
  \\[1.5mm]
  \,$X_\mu^a$~  & ~~$0$ \quad\quad  & $0$ & $0$ & $1$~~~
  \\[-2mm]
  &&&&
  \\ \hline \hline
 \end{tabular}
 \label{charge}
 \end{center}
 \end{table}

With proper choice of parameters, the singlet scalar $\sigma$ is
expected to develop a vacuum expectation value of the order of
TeV\,\footnote{It is clear
that, given $\,\left<\sigma\right>={\cal O}(\textrm{TeV})\,$ and
without fine-tuning the parameters, the mass of the physical Higgs
boson naturally lies at the TeV scale.} to drive the spontaneous
symmetry breaking of the global lepton number.
The vacuum expectation value $\,\left<\sigma\right>\,$ will induce
a small VEV for the singlet $\chi$ in the presence of the large mass
term and the trilinear coupling, as shown in the second line of
Eq.\,(\ref{potential}),
\begin{eqnarray}
\label{vev1}
\langle\chi\rangle &\simeq&
\frac{~\mu\langle\sigma\rangle^{2}_{}\,}{M_{\chi}^2} \,.
\end{eqnarray}
It is evident that the vacuum expectation value
$\langle\chi\rangle$ is highly suppressed
for $\,M_{\chi}^{}\gtrsim \mu \gg \langle\sigma\rangle\,$.\, This
clearly shares the essential feature with the traditional seesaw
mechanism. The small vacuum expectation value
$\langle\chi\rangle$ will then induce a
suppressed coupling of the Higgs triplet $\xi$ to the Higgs doublet
$\phi$ from the last terms of Eq.\,(\ref{potential}). According to
Eq.\,(\ref{trilinear}), we thus have,
\begin{eqnarray}
\label{trilinear-coupling}
\mu_0^{} ~=~ \kappa\langle\chi\rangle
~=~
\kappa\frac{\,\mu\langle\sigma\rangle^{2}_{}\,}{M_{\chi}^{2}}\,,
\end{eqnarray}
which is naturally small for \,$\kappa\lesssim{\cal O}(1)$\,.

\begin{figure}
\vspace{6.5cm} \epsfig{file=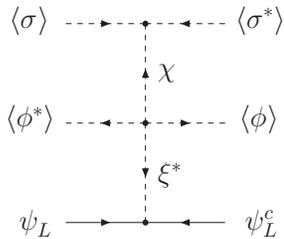, bbllx=5.3cm,
bblly=5.9cm, bburx=15.3cm, bbury=15.9cm, width=7.5cm, height=7.5cm,
angle=0, clip=0} \vspace{-10cm} \caption{\label{double-type-ii} The
new double type-II seesaw for the generation of Majorana neutrino
masses.} \vspace*{5mm}
\end{figure}

Subsequently, the Higgs doublet $\phi$ develops a VEV,
$\,\langle\phi\rangle\simeq 174\,$GeV, to break the
electroweak gauge symmetry \,$SU(2)_L\otimes U(1)_Y$.\,
So the Higgs triplet $\xi$ can pick up a small
VEV through the type-II seesaw mechanism,
\begin{eqnarray}
\label{vev2} \langle\xi\rangle &\simeq&
\frac{\,\mu_0^{}\langle\phi\rangle^{2}_{}\,}{m_{\xi}^2}\,.
\end{eqnarray}
We note that the VEV $\langle\xi\rangle$ is naturally small even for a
weak-scale triplet mass
$\,m_\xi^{}={\cal O}(\textrm{TeV})\gtrsim
 {\cal O}(\langle\phi\rangle)\,$,\,
thanks to the highly suppressed trilinear coupling $\mu_0^{}$ in
Eq.\,(\ref{trilinear-coupling}) through the tiny VEV
$\langle\chi\rangle$ in Eq.\,(\ref{vev1}). At this stage, the
neutrinos eventually obtain tiny Majorana masses through their
Yukawa couplings with the Higgs triplet as shown in
Eq.\,(\ref{yukawa}),
\begin{eqnarray}
\label{neutrino-masses}
m_{\nu}^{} &=& f\langle\xi\rangle\,.
\end{eqnarray}
Therefore, the small neutrino masses $\,m_{\nu}^{}\lll
\langle\phi\rangle\,$ can be naturally realized for the Yukawa
couplings $\,f={\cal O}(1)\,$.\, Remarkably, our new construction
includes an additional seesaw step to realize the type-II seesaw
mechanism as shown in Fig.\,\ref{double-type-ii}, and may thus be
called ``double type-II seesaw'' to reflect this new
ingredient\,\footnote{We can replace the heavy singlets by heavy
triplets to give another double type-II seesaw model,
\begin{eqnarray}
V &\supset&
-m_{1}^{2}(\sigma^{\dagger}_{}\sigma)
+\lambda_{1}^{}(\sigma^{\dagger}_{}\sigma)^{2}_{}
-m_{2}^{2}(\phi^{\dagger}_{}\phi)
+\lambda_{2}^{}(\phi^{\dagger}_{}\phi)^{2}_{}
\nonumber\\
&& +M_\chi^2\textrm{Tr}(\chi^\dagger_{}\chi)
-\left[\mu\sigma\textrm{Tr}(\xi^\dagger_{}\chi)
+\textrm{h.c.}\right]
\nonumber\\
&&
+m_\xi^2\textrm{Tr}(\xi^\dagger_{}\xi)
-(\kappa\sigma^\dagger_{}\phi^{T}_{}i\tau_2^{}\chi\phi+\textrm{h.c.})\,.
\end{eqnarray}
Here $\chi$ denotes the heavy triplets with the lepton number
$L=-1$, other notations coincide with the present model.
This model also has some similarity with the lepton number soft-breaking
models \cite{mss2008}.}.

For illustration, we give a numerical sample here. Let us take the
typical inputs, ~$M_{\chi}^{}=10^{14}\,\textrm{GeV}$,\, $\mu
=3\times 10^{12}\,\textrm{GeV}$, $\langle\sigma\rangle=
1.2\,\textrm{TeV}$, $m_{\xi}^{}=0.5\,\textrm{TeV}$, and
$\kappa=1$\,.\, We thus deduce, $\,\mu_0^{} = \langle\chi\rangle
\simeq 0.43\,\textrm{eV}$,\, and
\begin{eqnarray}
\label{example}
\langle\xi\rangle \simeq 0.05\,\textrm{eV}\,.
\end{eqnarray}
With the Yukawa coupling $\,f={\cal O}(1)\,$, we see from
(\ref{neutrino-masses}) that this naturally generates the small
neutrino masses,
$\,m_{\nu}^{} = f\langle\xi\rangle = {\cal O}(0.1)\,$eV,
consistent with the oscillation data \cite{ms2006}.

In the present construction, the cubic coupling (\ref{trilinear}) between
the triplet and doublet Higgs scalars is highly suppressed so that the
Higgs triplet $\xi$ can lie at the weak scale, with a mass around
$\mathcal{O}(0.5-1)$\,TeV, although its Yukawa couplings to the
leptons are of $\mathcal{O}(1)$.
Such a weak scale Higgs triplet will thus lead to interesting
phenomenology at the LHC and future lepton colliders (such as the
ILC\,\cite{ILC} or CLIC\,\cite{CLIC}). This Higgs triplet contains
neutral and charged Higgs bosons $(\xi^0,\,\xi^\pm,\,\xi^{\pm\pm})$.
At the LHC, they can be produced via their gauge couplings with
photon or $W^\pm/Z^0$ bosons and then decay {\it dominantly into
di-leptons.} For instance, these triplet Higgs bosons (with masses
around \,$0.5-1$\,TeV) can be detected at the LHC via $s$-channel
pair productions, \beqs
\begin{eqnarray}
\label{xi++xi--} && pp \to (\gamma^\ast ,Z^\ast) \to
\xi^{++}\xi^{--} \to \ell^+\ell^+\ell'^-\ell'^-\,,~~~~
\\[2mm]
\label{xi0xi0} && pp \to (Z^\ast)\to \xi^0\xi^0 \to
\ell^-\ell^+\ell'^-\ell'^+\,,~~~~
\\[2mm]
\label{xi++xi-} && pp \to (W^{+\ast})\to \xi^{++}\xi^- \to
\ell^+\ell^+\ell'^-\bar{\nu}'\,,~~~~
\\[2mm]
\label{xi+xi0} && pp \to (W^{+\ast})\to \xi^{+}\xi^0 \to
\nu\ell^+\ell'^-\ell'^+\,,~~~~
\end{eqnarray}
\eeqs
giving rise to distinct 4-lepton signatures or 3-lepton plus missing
energy signals. In particular, from (\ref{xi++xi--}) we see that the
like-sign 4-leptons of the type $\,e^\pm e^\pm \mu^\mp\mu^\mp$\,
with large invariant masses ($M_{ee}$ and $M_{\mu\mu}$) and large
transverse momenta are striking signals for the LHC discovery.
Another interesting process is the pair production,
\begin{eqnarray}
\,pp\to
jj\xi^{++}\xi^0\to jj\ell^+\ell^+\ell'^+\ell'^-\,,
\end{eqnarray}
via the quartic vertex $\,W^\pm W^\pm \xi^{\mp\mp}\xi^0$ or the $t$-channel
$\xi$-exchange,  with two energetic forward-jets and 4-leptons (such as
$e^+e^+\mu^+\mu^-$ and $\mu^+\mu^+e^+e^-$) in the final state.
Furthermore, at a future TeV lepton collider (such as ILC or CLIC)
operating in the $e^-e^-$ mode,
the $\xi^{\pm\pm}$ can be directly produced via $\,e^-e^-\to
\xi^{--}\to  \ell^-\ell^-\,$, with $\ell^-\ell^- = \mu^-\mu^-$ for
instance, which has very clean background.\footnote{If such
double-charged Higgs bosons $\xi^{\pm\pm}$ fall in the mass-range of
$300-500$\,GeV as could be seen at the LHC \cite{aa2005etc},
a 500\,GeV ILC operated in the $e^-e^-$ mode is expected to
confirm the $\xi^{--}$ discovery
via the $s$-channel production, and further measure
its properties precisely.}

We also note that since lepton number is a global symmetry, its
spontaneous breaking induced by the nonzero $\langle\sigma\rangle$
will lead to a massless Nambu-Goldstone boson. Although this
Nambu-Goldstone boson has a component from the Higgs triplet, there
is no problem with the low-energy phenomenology including the LEP
constraints because this triplet fraction is highly suppressed by
\,$\langle\xi\rangle/\langle\sigma\rangle \sim
\mathcal{O}(10^{-13})$.\, Also, the imaginary parts of the Higgs
triplet $\xi$ and the doublet $\phi$ have mixing,
and one of their combinations results in a physical pseudoscalar
which has a mass controlled by the heavier mass \,$m_\xi$\, and
thus escapes from the low-energy constraints.

\section{III. Baryogenesis via Leptogenesis}

\begin{figure*}
\label{leptogenesis}
\vspace{4.5cm}
\epsfig{file=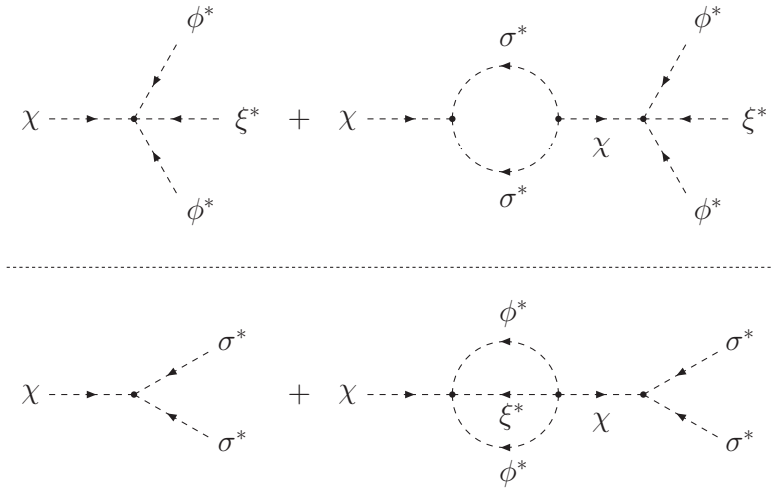, bbllx=5.3cm,
bblly=5.9cm, bburx=15.3cm, bbury=15.9cm, width=8cm, height=8cm,
angle=0, clip=0} \vspace{-5.4cm}
\caption{The lepton number conserving decays of the heavy singlets $\chi$
at the tree level and the loop orders for generating a lepton asymmetry
stored in the Higgs triplet $\xi$\,,\, and an equal but opposite amount
stored in the light singlet $\sigma$\,.\, Here the CP-conjugate diagrams
are not shown for simplicity.}
\end{figure*}

In the present model, the heavy singlets $\chi$ have two decay channels as
shown in Fig.\,\ref{leptogenesis},
\begin{eqnarray}
\label{eq:chi-decays}
\chi\,\rightarrow\, \xi^\ast_{}\,\phi^\ast_{}\,\phi^\ast_{} \quad
\textrm{and}\quad \chi\,\rightarrow\, \sigma^\ast_{}\,\sigma^\ast_{}
\,.
\end{eqnarray}
If CP is not conserved, the $\,\chi \rightarrow
\xi^\ast_{}\phi^\ast_{}\phi^\ast_{}\,$ process and its CP-conjugate
can generate a lepton asymmetry stored in the Higgs triplet after
they go out of equilibrium. At the same time, there will emerge an
equal but opposite lepton asymmetry stored in the light singlet
since the lepton number is conserved in the $\chi$ decays and the
sum of lepton asymmetries from the two decay channels in
(\ref{eq:chi-decays}) vanishes.  Note that in the leptogenesis,
the sphaleron action has no effect on the light singlet. So we
can focus on the lepton asymmetry stored in the Higgs triplet, which
has been decoupled from the lepton asymmetry stored in the light
singlet and will be rapidly transferred to the lepton doublets.
After the lepton number is spontaneously broken at the TeV scale by
the VEV of the singlet scalar $\sigma$, the Higgs triplet $\xi$ will
develop a trilinear coupling $\mu_0^{}$ to the Higgs doublet $\phi$,
as shown in Eqs.\,(\ref{trilinear}) and (\ref{trilinear-coupling}).
Due to the smallness of this trilinear coupling $\mu_0^{}$, the
induced lepton number violating processes take place so slowly that
they will not reach equilibrium until the temperature falls well
below the electroweak scale, where the sphaleron process has become
very weak. In consequence, the lepton asymmetry stored in the Higgs
triplet can be partially converted to a baryon asymmetry.

It is clear that at least two such heavy singlets are needed to have
CP violation and realize an interference between the tree-level
diagram and the loop-order self-energy. Here we minimally introduce
two heavy singlets $\,\chi_{1,2}^{}$\,.\, For convenience, we choose a
basis of the heavy singlets by a proper rotation, which gives real,
diagonal scalar mass-matrix \,$M_{\chi}^{2}
=\textrm{diag}(M_{\chi_1^{}}^{2},\, M_{\chi_2^{}}^{2})$\, and two real
cubic scalar-couplings $\,\mu=(\mu_{1}^{},\,\mu_{2}^{})$.\,
Consequently, we only need to keep
$\,\kappa=(\kappa_1^{},\,\kappa_2^{})$\, complex in the Higgs
potential. For illustration, consider the case where the two heavy
singlets $\chi_{1,2}^{}$ have hierarchical mass-spectrum. In this
case, the final lepton asymmetry stored in the Higgs triplet will
mainly come from the decay of the lighter one.  Without the loss of
generality, we choose $\chi_{1}^{}$ to be the lighter singlet and
$\chi_{2}^{}$ the heavier one, i.e., \,$M_{\chi_1}\ll M_{\chi_2}$.\,
We then compute the CP-asymmetry for the Higgs triplet,
\begin{eqnarray}
\varepsilon_{1}^{}
&=& 2\frac{\Gamma(\chi_{1}^{} \rightarrow
    \xi^\ast_{}\phi^\ast_{}\phi^\ast_{})
    -\Gamma(\chi_{1}^{\ast}\rightarrow\xi\phi\phi)}{\Gamma_{1}^{}}
\nonumber\\[3mm]
&\simeq& \frac{1}{2\pi}
\frac{\textrm{Im}(\kappa_{1}^{\ast}\kappa_{2}^{})}{|\kappa_{1}^{}|^{2}_{}}
\frac{\mu_{1}^{}\mu_{2}^{}}{M_{\chi_2^{}}^{2}-M_{\chi_1^{}}^{2}}
\frac{\frac{3}{32\pi^{2}_{}}|\kappa_{1}^{}|^{2}_{}}
{\frac{\mu_{1}^{2}}{M_{\chi_{1}^{}}^{2}}
+\frac{3}{32\pi^{2}_{}}|\kappa_{1}^{}|^{2}_{}}
\nonumber %\\[3mm]
\end{eqnarray}
\begin{eqnarray}
&=&
\frac{\sin\delta}{2\pi}\left|\frac{\kappa_{2}^{}}{\kappa_{1}^{}}\right|
\frac{\mu_{1}^{}\mu_{2}^{}}{M_{\chi_2^{}}^{2}-M_{\chi_1^{}}^{2}}
\frac{\frac{3}{32\pi^{2}_{}}|\kappa_{1}^{}|^{2}_{}}
{\frac{\mu_{1}^{2}}{M_{\chi_{1}^{}}^{2}}
+\frac{3}{32\pi^{2}_{}}|\kappa_{1}^{}|^{2}_{}}\,,
\label{eq:epsilon-1}
\end{eqnarray}
where we have defined $\,\kappa_i = |\kappa_i|e^{i\delta_i}\,$ and
$\,\delta \equiv \delta_2 - \delta_1$\, is the difference of the two
phase angles which controls the above CP-asymmetry parameter
$\varepsilon_{1}^{}$.\, In (\ref{eq:epsilon-1}), $\Gamma_{i}^{}$
denotes the total decay width of $\chi_{i}^{}$ or $\chi_{i}^{\ast}$,
\begin{eqnarray}
\label{ucpt} \Gamma_{i}^{}&=&\Gamma(\chi_{i}^{\,\,\,}\rightarrow
\xi^\ast_{}\phi^\ast_{}\phi^\ast_{})+\Gamma(\chi_{i}^{}\,
\rightarrow \sigma^\ast_{}\sigma^\ast_{})
\nonumber\\[2mm]
&=&\Gamma(\chi_{i}^{\ast}\rightarrow\xi^{\,\,\,}_{}
\phi^{\,\,\,}_{}\phi^{\,\,\,}_{})+\Gamma(\chi_{i}^{\ast} \rightarrow
\sigma^{\,\,\,}_{}\sigma^{\,\,\,}_{})
\nonumber\\[3mm]
 &=&\frac{1}{8\pi}\left(\frac{\mu_{i}^{2}}{M_{\chi_{i}^{}}^{2}}
 +\frac{3}{32\pi^{2}_{}}|\kappa_{i}^{}|^{2}_{}\right)M_{\chi_{i}^{}}^{}\,.
\end{eqnarray}
Here the second equality is guaranteed by the unitarity and the CPT
conservation.

When the lepton number is spontaneously broken by the VEV
$\langle\sigma\rangle$, there will emerge a trilinear coupling of
the Higgs triplet to the Higgs doublet as shown in
Eqs.\,(\ref{trilinear}) and (\ref{trilinear-coupling}). For
$\langle\sigma\rangle=1.2\,\textrm{TeV}$, the phase transition may
occur at the temperature $\,T_{c}^{}\,\lesssim\,
m_{\xi}^{}=0.5\,\textrm{TeV}$,\, where the lepton number violating
processes have been already decoupled because
\begin{eqnarray}
\Gamma(\xi\!\rightarrow\! \phi^\ast_{}\phi^\ast_{}) \,=\,
\f{1}{16\pi}\frac{|\mu_0^{}|^{2}_{}}{m_{\xi}^{}} &\ll&
H(T)\!\left|^{}_{T\simeq m_{\xi}^{}}\right. \!.
\end{eqnarray}
Here,
\begin{eqnarray}
\label{hubble}
H(T)&=&\left(\frac{8\pi^{3}_{}g_{\ast}^{}}{90}\right)^{\frac{1}{2}}_{}
\frac{T^{2}_{}}{M_{\textrm{Pl}}^{}}
\end{eqnarray}
is the Hubble constant, with the relativistic degrees of freedom
\,$g_{\ast}^{}\simeq 100$\, and the Planck mass
$M_{\textrm{Pl}}^{}\simeq 10^{19}_{}\,\textrm{GeV}$. We then
derive the final baryon asymmetry by computing the ratio of the
baryon number density $n_B^{}$ over the entropy density $s$\,,
\begin{eqnarray}
\label{baryon-asymmetry} \frac{n_B^{}}{s} &=& \frac{28}{79}\,
\frac{n_{B}^{}-n_{L_{SM}^{}}^{}}{s} ~=~
-\frac{28}{79}\,\frac{n_{L_{SM}^{}}^{}}{s}
\nonumber\\[3mm]
&\simeq& -\left.
\frac{28}{79}\,\varepsilon_{1}^{}\frac{n_{\chi_1^{}}^{eq}}{s}\right|_{T\simeq
M_{\chi_1^{}}^{}}^{} \simeq~
-\frac{1}{15}\frac{\varepsilon_{1}^{}}{g_\ast^{}}\,,
\end{eqnarray}
where \,$n_{\chi_1^{}}^{eq}$\, is the thermal equilibrium density of
$\,\chi_{1}^{}\,$.\,  Note the above solution is only valid for the weak
washout regime with
\begin{eqnarray} \label{equilibrium}
\Gamma_{\chi_1^{}}^{} &\lesssim & H(T)\left|_{T\simeq
M_{\chi_1}^{}}^{}\right.\,.
\end{eqnarray}
Let us consider the sample inputs,
\begin{eqnarray}
&&
M_{\chi_1^{}}^{}=0.1\,M_{\chi_2^{}}^{} =
10^{14}_{}\,\textrm{GeV}\,, \nonumber\\[3mm]
&&
\mu_{1}^{} =\mu_{2}^{}=3\times10^{12}_{}\,\textrm{GeV}\,,
\\[3mm]
&& |\kappa_{1}^{}|=|\kappa_2^{}|=1\,, ~~~
 \sin\delta=-0.11 \,,
\nonumber
\end{eqnarray}
with which we can estimate the CP-asymmetry,
\begin{eqnarray}
\label{eq:CPAS} \varepsilon_{1}^{} &\simeq& -1.3\times 10^{-7}_{}\,.
\end{eqnarray}
For convenience, we express the baryon asymmetry in terms of the
ratio of $n_B^{}$ over the photon density $n_\gamma^{}$\,,
\begin{eqnarray}
\label{eq:nB-nGamma} \f{n_B^{}}{n_\gamma^{}} &=& 7.04\f{n_B^{}}{s}
~\simeq~ 6.3\times 10^{-10} \,,
\end{eqnarray}
which is consistent with the five-year observations of the WMAP
collaboration \cite{dunkley2008}, $\,n_B^{}/n_\gamma^{}=(6.225\pm
0.170)\times 10^{-10}\,$.

In the present model, the leptogenesis mechanism is different from
that in the conventional seesaw models. Now the amount of lepton asymmetry
does not depend on the parameters in the neutrino mass matrix.
As a result, there is no DI bound \cite{di2002} on the
decaying particles. So we are flexible to lower the leptogensis scale
and avoid the gravitino problem in a supersymmetric extension of this
model, although we do not elaborate on this point
in the present paper.

\section{IV. Leptonic dark matter}

The recent PAMELA experiment\,\cite{adriani2008} found an anomalous
rise of the $e^+/(e^++e^-)$ fraction in cosmic rays, while the
HESS\,\cite{aharonian2008} and Fermi/LAT\,\cite{fermi2009}
observations further exhibit an excess over the conventional
background-predictions for the cosmic ray fluxes. Systematical data
fitting shows\,\cite{strumia0905} that for the dark matter particles
of mass around 3\,TeV, their annihilations into the leptonic final
states ($\mu^+_{}\mu^-_{},\tau^+_{}\tau^-_{},4\mu,4\tau$, depending
on the dark matter profile) can give a good fit to the excess; in
general, a dark matter having mass lighter than about a TeV is
excluded. In this section, we will extend our model to include a
dark matter sector, in which the vector dark-matter particles
annihilate dominantly into the $\mu$'s and $\tau$'s. With the aid of
Breit-Wigner enhancement and for a hidden vector boson mass around
3\,TeV, we find that our model can naturally explain the cosmic
$e^\pm$ excesses based on the general analysis in
\cite{strumia0905}.

To construct the dark matter sector, we consider a hidden $SU(2)_h$ gauge
theory with a complex Higgs doublet field \,$\eta$\,,\,  but
with no extra fermions (cf.\ Table\,1). So, the
hidden sector Lagrangian for $\eta$ doublet can be written as,
\begin{eqnarray}
\label{dark}
\mathcal{L}_{\eta} &=&
\left(D_{\mu}^{}\eta\right)^\dagger_{}\left(D^{\mu}_{}\eta\right)-V(\eta)\,,
\end{eqnarray}
where
\begin{eqnarray}
D_{\mu}^{}\eta &=&\left( \partial_{\mu}^{} -
ig_{X}^{}\frac{\vec{\tau}}{2}\cdot \vec{X}_{\mu}^{}\right)\eta\,,
\\[2mm]
V(\eta)&=& -m_{3}^{2}\left(\eta^\dagger_{}\eta \right)+
\lambda_{3}^{} \left(\eta^\dagger_{}\eta\right)^{2}_{}\,.
\end{eqnarray}
After the hidden $SU(2)_h$ is spontaneously broken by the vacuum expectation
value \,$\left<\eta\right>$\,,
we are left with three degenerate massive vector bosons $\vec X_\mu$
as well as one neutral physical Higgs boson \,$\zeta$\,,
\begin{eqnarray}
\mathcal{L}_\eta
& \supset &
\frac{1}{2}m_{X}^{2}|\vec{X}_{\mu}^{}|^2
+\frac{1}{2}m_{\zeta}^{2}\zeta^2_{}
\nonumber\\[1.5mm]
&&+\frac{1}{8}g_{X}^{2}\zeta^{2}_{} |\vec{X}_{\mu}^{}|^2
 +\frac{1}{2\sqrt{2}}g_{X}^{2}
  \langle\eta\rangle\zeta|\vec{X}_{\mu}^{}|^2 \,,
  \label{eq:L-eta}
\end{eqnarray}
with
\begin{eqnarray}
m_{X} ~=~\frac{1}{\sqrt 2}g_{X}^{}\langle\eta\rangle \,,~~~~
m_{\zeta} ~=~ \sqrt{2}m_{3}^{}\,.
\end{eqnarray}
The hidden $SU(2)_h$ forbids its associated vector bosons to mix
with the SM gauge bosons because of its non-Abelian character.
Consequently the hidden vector bosons are stable, with no decay
channel.

\begin{figure}
\vspace{0.5cm} \epsfig{file=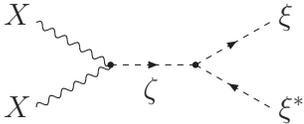, bbllx=4.3cm,
bblly=5.9cm, bburx=14.3cm, bbury=15.9cm, width=8cm, height=8cm,
angle=0, clip=0} \vspace{-5.7cm} \caption{\label{dark-matter}
The hidden vector bosons annihilate into Higgs triplets through the
$s$-channel exchange of the hidden physical Higgs boson.}
\end{figure}

The hidden sector can communicate with the visible sector through
the following quartic scalar interactions,
\begin{eqnarray}
\label{eq:alpha-beta-terms}
\mathcal{L}
&\supset&
-\alpha\left(\eta^\dagger_{}\eta\right)
       \textrm{Tr}\left(\xi^\dagger_{}\xi\right)
-\beta\left(\eta^\dagger_{}\eta\right)
      \left(\phi^\dagger_{}\phi\right) ,~~~
\end{eqnarray}
where the $\alpha$-term involves the interaction between the hidden
Higgs doublet $\eta$ and the Higgs triplet $\xi$\,,\, while the
$\beta$-term links $\eta$ to the SM Higgs doublet $\phi$\,.\, Once
$\eta$ develops a vacuum expectation value $\left<\eta\right>$, the
$\alpha$-term and $\beta$-term will induce the cubic scalar
couplings for \,$\eta-\xi-\xi$\, and $\,\eta-\phi-\phi\,$ vertices,
respectively. In the unitary gauge, the Higgs doublet $\eta$ (or
$\phi$) reduces to the neutral physical Higgs boson $\zeta^0$ (or
$h^0$). Because of the $\,\zeta-X^a_\mu-X^{a\mu}\,$ cubic
interaction in (\ref{eq:L-eta}), we have the dark matter
annihilation processes\,\footnote{We may also have decay processes
for non-thermal production of dark matter\,\cite{lhzb2001}. For
instance, the heavy singlets $\chi$ may decay into the hidden vector
bosons $X$ due to the quartic interaction between $\chi$ and the
hidden Higgs doublet $\eta$,
$\,\left(\chi^\dagger_{}\chi\right)\left(\eta^\dagger_{}\eta\right)$\,.\,
However, this decay is highly suppressed because of the tiny VEV
$\langle\chi\rangle$ in (\ref{vev1}) and the heavy mass $M_\chi^{}$
in (\ref{potential}).\, So, the induced relic density of dark matter
is too small to be relevant.}, $\,X^aX^a \to \xi\xi^\ast\,$ and
$\,X^aX^a \to hh\,$,\, with the $s$-channel $\zeta$-exchange, where
$\,\xi \in (\xi^0,\,\xi^\pm,\,\xi^{\pm\pm})\,$ contains $6$ real
degrees of freedom. For the scattering energy much larger than the
masses of $\xi$ and $h$, we find that the ratio of the inclusive
cross sections for the final states $\,\xi\xi^\ast\,$ and $\,hh\,$
is about $\,6(\alpha/\beta)^2\,$.\, Let us consider the natural
parameter space of $\,\beta \leqslant (0.3-0.5)\alpha\,$ without any
fine-tuning. We can thus estimate the ratio of the two inclusive
cross sections, \beqa \label{eq:ratio} \f{\sigma[X^aX^a\to
hh]}{\sigma[X^aX^a\to \xi\xi^\ast]} &\simeq&
\f{1}{6}\left(\f{\beta}{\alpha}\right)^2 ~\leqslant~ 0.5\%-4\%
\,,~~~~~~ \eeqa indicating that the process $\,X^aX^a \to
\xi\xi^\ast\,$ fully dominates the dark matter annihilation cross section.
Due to the large triplet Yukawa couplings \,$f={\cal O}(1)$\, in (\ref{yukawa})
and tiny trilinear scalar coupling
\,$\mu_0^{}={\cal O}(0.1-1)$eV\, in (\ref{trilinear}),
our dark matter annihilation $\,X^aX^a \to \xi\xi^\ast\,$ will naturally lead to
leptonic decay products\,\footnote{As variations of our current
construction, we may also consider the dark matter annihilation into
other leptonic scalars like those in Zee model\,\cite{zee1980} or
Zee-Babu model\,\cite{zee1985}.} and thus nicely agree with the
exciting signals from the recent cosmic ray
experiments\,\cite{chang2008,torii2008,adriani2008,aharonian2008,fermi2009},
as mentioned at the beginning of this section. On the other hand,
for the process $\,X^aX^a \to hh\,$, the SM Higgs boson $h$ will
decay preferably into $b\bar{b}$ or $WW/ZZ$ final states, but will
not cause visible signals in the present cosmic ray data due to its
too small cross section in (\ref{eq:ratio}).

So, we consider the hidden vector boson annihilation
into the Higgs triplets,\footnote{As the present work is being prepared,
a new preprint\,\cite{gos2009} considered scalar dark matter with their
annihilations into Higgs triplets.}\,
$\,X^a_{}X^a_{}\to \xi\xi^\ast~(a=1,2,3)$,\, through the
$s$-channel exchange of the hidden Higgs boson $\zeta$\,,\, as shown
in Fig.\,\ref{dark-matter}.
The form of the cubic scalar vertex \,$\zeta-\xi-\xi$\, can be derived
from the $\alpha$-term in (\ref{eq:alpha-beta-terms}),
\begin{eqnarray}
\mathcal{L}&\supset& -\alpha\left(\eta^\dagger_{}\eta\right)
\textrm{Tr}\left(\xi^\dagger_{}\xi\right)
\nonumber\\[2mm]
&\supset&
-\sqrt{2}\alpha\langle\eta\rangle\,\zeta\,
\textrm{Tr}\left(\xi^\dagger_{}\xi\right)\,,
\end{eqnarray}
To compute the cross section of this process, we will average over the
initial state polarizations and gauge indices.
Thus, we can derive the unpolarized cross section for
$\,X^a_{}X^a_{}\to \xi\xi^\ast$\,,
\begin{eqnarray}
\label{cross-section}
\sigma v & =&
\frac{\,g_{X}^{4}\alpha^{2}_{}\,}{72\pi s}
\left[2+\frac{\left(s-2m_{X}^2\right)^{2}_{}}{4m_{X}^4}\right]
\frac{\langle\eta\rangle^{4}_{}}{(s-m_{\zeta}^{2})^2_{}
+m^2_{\zeta}\Gamma^2_{\zeta}} \,,
\nn\\
\end{eqnarray}
where $v$ is the relative velocity of the initial state vector bosons,
\beqa
v &=& 2\sqrt{1-\f{4m_X^2}{s}\,}\,,
\eeqa
while the decay width of the Higgs boson $\zeta$ is given by
the channel $\,\zeta\to \xi\xi^\ast$\,
(for $\,m_\zeta^{}<2m_X^{}$),
\begin{eqnarray}
\Gamma_{\zeta}^{}&=&
\frac{\,3\alpha^{2}_{}\,}{8\pi}
\frac{\langle\eta\rangle^{2}_{}}{m_{\zeta}^{}\,} \,.
\end{eqnarray}

We now verify that the hidden vector bosons can indeed serve as the
dark matter. For this purpose, we need to thermally average the
cross section (\ref{cross-section}) and determine their relic
density. Following Refs.\,\cite{RS-GS,ckrs2008,imy2008,gw2009}, we
consider the Breit-Wigner resonant case,
\begin{eqnarray}
\delta &=& 1-\frac{m_{\zeta}^{2}}{4m_{X}^{2}} ~\ll~ 1\,.
\end{eqnarray}
The relic density of the dark matter in our model
can be solved from the Boltzmann equation\,\cite{Kolb-Turner}
with the resonant effect\,\cite{imy2008},
\begin{eqnarray}
\label{eq:Omegah2}
\Omega_{\textrm{DM}}^{}h^2_{} &\simeq&
\frac{1.07\times 10^9x_f\,\textrm{GeV}^{-1}
}{\sqrt{g_\ast^{}}\,M_{\rm Pl}\langle\sigma v\rangle_{0}^{}}\times \textrm{BF}
\nonumber\\[3mm]
&\simeq&
\frac{\,0.1\times 10^{-9}\textrm{GeV}^{-2}\,}{\langle\sigma v\rangle_{0}^{}}
\f{\,x_f^{}}{\sqrt{g_\ast^{}}\,}\times \textrm{BF}\,,
\end{eqnarray}
where BF is the
effective boost factor due to Breit-Wigner
resonant effect\,\cite{imy2008}, as will be discussed shortly.  The parameter
$\,x_f^{}=m_{\rm DM}^{}/T\,$ is the freeze-out temperature\,\cite{Kolb-Turner},
\beqs
\beqa
x_f^{} &=& \ln X -\f{1}{2}\ln\ln X \,,
\\[3mm]
X &=& 0.038(g/\sqrt{g_\ast^{}})M_{\rm Pl}\left<\sigma v\right>_{0} \,,
\eeqa
\eeqs
with $g_\ast^{}\,(g)$ the degrees of freedom for massless particles (dark matter).
Here we have $\,g_\ast^{}\simeq 100\,$ and $\,g=3\times 3=9\,$ in the present model.
The $x_f^{}$ is typically of \,${\cal O}(10-30)$.\,
In the above we have defined
$\,\left<\sigma v\right>_0 =\left<\sigma v\right>|_{T=0}^{}\,$.\,
For the $s$-wave dark matter annihilation in the non-relativistic limit, the thermally
averaged cross section equals the non-averaged one,
$\,\left<\sigma v\right>\simeq\sigma v\,$,\, so we can deduce from (\ref{cross-section}),
\beqs
\begin{eqnarray}
\langle\sigma v\rangle_{0}^{} &\simeq&
\frac{g_{X}^{2}}{\,72m_{X}^{2}\,}
\frac{\gamma}{\delta^{2}_{}+\gamma^2_{}}\,,
\\[3mm]
\gamma &=&
\frac{\Gamma_{\zeta}^{}}{m_{\zeta}^{}}
~=~ \frac{3\alpha^2}{16\pi g_X^2}\frac{1}{1-\delta}\,.
\end{eqnarray}
\eeqs
where, besides $\,\delta \ll 1\,$, we also note $\,\gamma\ll 1\,$
for $\,\alpha /g_X^{} < 0.1$\,.\,
Finally, the effective boost factor BF due to Breit-Wigner
resonant effect is given by\,\cite{imy2008},
\beqa \textrm{BF} &\simeq&
\frac{\max[\delta,\gamma]^{-1}_{}}{x_f^{}}\,, \eeqa which is
expected to be around ${\cal }\mathcal{O}(10^2-10^3)$\,.
The effective boost factor for the annihilation cross section
may alternatively arise from Sommerfeld enhancement\,\cite{Sommerfeld},
which will not be explored in the present study.

\begin{widetext}
\begin{figure*}
\vspace*{3mm}
  \centering
  \includegraphics[width=10cm]{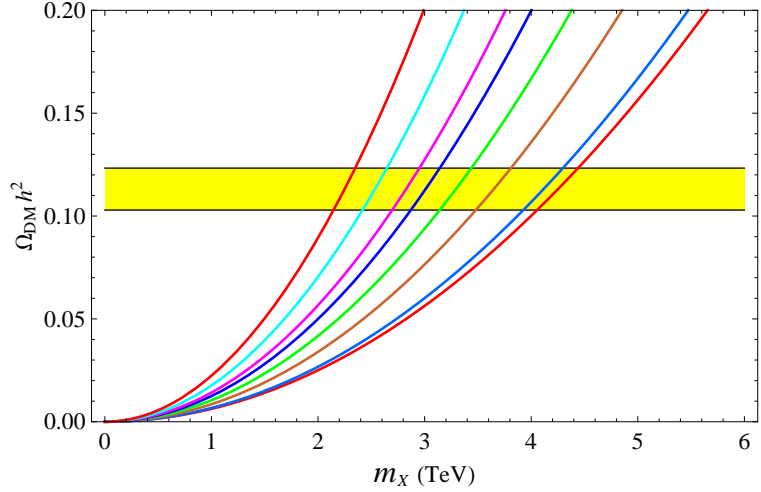}
  \vspace*{-2mm}
  \caption{Predictions of the dark matter density $\Omega_{\rm DM}^{} h^2$ as a function of
  hidden vector boson mass $m_X^{}$ in our model. The curves from right to left
  correspond to  $\,g_X^{}/\alpha=10,\,20,\,30,\,35,\,40,\,50,\,60,\,70$, respectively,
  where we have set $\,\delta=0.4\times10^{-4}_{}$ and $\,g_X^{}=0.34\,$.
  The shaded region (yellow) is allowed by the cosmological bound,
  $\,0.1029 < \Omega_{\rm DM}^{} h^2 < 0.1233$ ($3\sigma$ level),
  as extracted from the WMAP data\,\cite{dunkley2008}.
  }
  \label{Fig4}
\end{figure*}
\end{widetext}

For illustration, let us consider a sample input,
\begin{eqnarray}
\label{parameter}
\delta = 0.4\times 10^{-4}_{},~~~
g_{X}^{}=40\alpha = 0.34\,,~~~
m_{X}^{} = 3\,\textrm{TeV}.~~~
\end{eqnarray}
With this we can thus derive,
\beqs
\begin{eqnarray}
&&
x_f^{}~\simeq~ 27\,, ~~~~
\gamma ~\simeq~ 0.37\times 10^{-4}_{}\,,
\\[2mm]
\label{BF-gamma}
&& \textrm{BF} ~\simeq~ 0.93\times 10^3\,,
\\[2mm]
\label{cross-section-2}
&&
\langle\sigma v\rangle_{0}^{} ~\simeq~
2.2\times 10^{-6}_{}\,\textrm{GeV}^{-2}_{}\,,
\end{eqnarray}
\eeqs
and the relic density for dark matter,
\begin{eqnarray}
\Omega_{\textrm{DM}}^{}h^2_{}
&\simeq& 0.11\,,
\end{eqnarray}
which is consistent with the measured value given by the WMAP
observations (combined with the distance measurements from Type-Ia
Supernovae and the Baryon Acoustic Oscillations in the
distributions of galaxies)\cite{dunkley2008},
$\,\Omega_{\textrm{DM}}^{}h^2_{} = 0.1131 \pm 0.0034\,$.\,

To further explore the parameter space, we
have plotted Fig.\,4, where the relic dark matter density
$\Omega_{\rm DM}^{} h^2$ is shown as a function of the hidden vector
boson mass $m_X^{}$ for possible values of the coupling-ratio
$\,g_X^{}/\alpha$\,.\, The present WAMP constraint, $\,0.1029 <
\Omega_{\rm DM}^{} h^2 < 0.1233$ ($3\sigma$ level), is imposed as
the shaded region in the parameter space.

In the galactic halo, the annihilating dark matter has a relative
velocity about $v\sim10^{-3}$, and its thermally averaged cross
section is well described by that at the zero temperature for
$v^{2}_{}\ll\delta,\gamma$. From the parameter choice
(\ref{parameter}), we see that \,$v^{2}_{}\ll \gamma < \delta $\,
holds. It is known that a very small $\delta$ does not appear technically
so natural \cite{RS-GS}-\cite{gw2009}, while
the Breit-Wigner enhancement can give the desired large annihilation cross section of
$\,{\cal O}(10^{-6})\,$GeV$^{-2}$\, as in (\ref{cross-section-2}),
and is consistent with the recent cosmic-ray signals
\cite{adriani2008,aharonian2008,fermi2009}.
In our model, the anomalies from the PAMELA\,\cite{adriani2008},
HESS\,\cite{aharonian2008} and Fermi/LAT\,\cite{fermi2009}
observations can be understood in two steps: (i) first, the three
degenerate vector bosons $X^a$ with mass around 3\,TeV, annihilate
into the Higgs triplet $\xi$ with the enhanced cross section
(\ref{cross-section-2}); (ii) subsequently, the Higgs triplet $\xi$
rapidly and mostly decays into the leptons (the $\mu$'s and
$\tau$'s).

\vspace{3mm}

\section{V. Summary}

We have constructed a new double type-II seesaw scenario where the
coupling of the Higgs triplet to the Higgs doublet is highly
suppressed after spontaneous breaking of the global lepton number.
In our model, the small neutrino masses can be naturally realized,
while at the same time the Yukawa couplings of the TeV-scale Higgs
triplet to the lepton doublets are large enough, leading exciting
phenomenology at the LHC and future TeV lepton colliders (such as
the ILC or CLIC). Furthermore, a lepton asymmetry stored in the
Higgs triplet can be generated in the lepton number conserving
decays and then be rapidly transferred to the lepton doublets.
Hence, the matter-antimatter asymmetry in the Universe can be
naturally explained via leptogenesis in our model. We have further
extended the model to realize the leptonic dark matter by including
stable vector bosons associated with a hidden non-Abelian gauge
symmetry. The stable vector bosons can dominantly annihilate into
the Higgs triplets, which decay mostly into leptons in our double
type-II seesaw scenario. In this annihilation, there is an
$s$-channel process mediated by a hidden Higgs boson, which is
responsible for the mass generation of dark matter. Furthermore, the
Breit-Wigner resonant enhancement makes it possible to have a large
annihilation rate of the dark matter particles into the $\mu$'s and
$\tau$'s and thus explain the cosmic $e^\pm$ excesses from our
model.

\vspace{5mm}

\noindent \textbf{Acknowledgement}: PHG thanks Goran
Senjanovi$\rm\acute{c}$ for helpful discussions. HJH is supported in
part by the National Natural Science Foundation of China under
grants 10625522 and 10635030. US thanks the Department of Physics
and the McDonnell Center for the Space Sciences at Washington
University in St.\ Louis for inviting him as Clark Way Harrison
visiting professor. XZ is supported in part by the National Natural
Sciences Foundation of China under grants 10533010 and 10675136, and
by the Chinese Academy of Sciences under grant KJCX3-SYW-N2.

\end{document}